\documentclass[preprint,eqsecnum,preprintnumbers,nofootinbib,byrevtex,prd,aps,showpacs,showkeys,groupedaddress,floatfix]{revtex4}
\usepackage{bm}
\usepackage{graphics}
\usepackage{graphicx}
\usepackage{epsfig}
\usepackage{amssymb}
\usepackage{amsmath}
\begin{document}

\title{HIGHER TWIST EFFECTS IN PHOTON-PHOTON COLLISIONS}
\author{A.~I.~Ahmadov$^{1,2}$} \email{E-mail:ahmadovazar@yahoo.com}
\author{I.~Boztosun$^{3}$}
\author{A.~Soylu$^{3}$}
\author{E.~A.~Dadashov$^{2}$}%
\affiliation{$^{1}$Physikalisches Institut, Universit\"{a}t Bonn,
Nussallee 12, D-53115, Bonn, Germany \\ $^{2}$ Institute for Physical Problems,
Baku State University, Z. Khalilov st. 23, AZ-1148, Baku, Azerbaijan \\
$^{3}$Department of Physics, Faculty of Arts and
Sciences, Erciyes University, Kayseri, Turkey }

\date{\today}
\begin{abstract} In this article, we investigate the contribution of
the high twist Feynman diagrams to the large-$p_T$ single
pseudoscalar and vector mesons inclusive production  cross section
in two-photon collisions and we present the general formulae for
the high and leading twist differential cross sections. The pion
wave function where two non-trivial Gegenbauer coefficients $a_2$
and $a_4$ have been extracted from the CLEO data, Braun-Filyanov
pion wave function, the asymptotic and the Chernyak-Zhitnitsky
wave functions are used in the calculations. For $\rho$-meson we
used Ball-Braun wave function. The results of all the calculations
reveal that the high twist cross sections, the ratio $R$,
the dependence transverse momentum $p_T$ and the rapidity $y$ of
meson in the $\Phi_{CLEO }(x,Q^2)$ wave function case is very close
to the $\Phi_{asy}(x)$ asymptotic wave function case. It is shown
that the high twist contribution to the cross section depends on
the choice of the meson wave functions.
\end{abstract}
 \maketitle

\section{\bf Introduction}

During the last few years, a great deal of progress has been made
in the investigation of the properties of hadronic wave
functions[1-12]. The notion of distribution amplitudes refers to momentum
fraction distributions of partons in meson in particular Fock state with fixed
number of compenents. For the minimal number of constituents, the distribution
amplitude $\Phi$ is related to the Bethe-Salpeter wave function $\Phi_{BS}$ by

\begin {equation}
\Phi(x)\sim\int^{|k_{\perp}|<{\mu}}d^{2}k_{\perp}\Phi_{BS}(x,k_{\perp}).
\end {equation}

The standard approach to distribution amplitudes, which is due to Brodsky and
Lepage[13], considers the hadron's parton decomposition in the infinite
momentum frame. A conceptually different, but mathematically equivalent
formalizm  is the light-cone quantization[14]. Either way, power suppressed
contributions to exclusive processes in QCD, which are commonly refered to as
higher twist corections. The meson wave functions (also called
distribution amplitudes -DA) [1] play a key role in the
hard-scattering QCD processes because they encapsulate the
essential nonperturbative features of the meson's internal
structure in terms of the parton's longitudinal momentum fractions
$x_i$. Meson wave functions have been extensively studied by using
QCD sum rules. The original suggestion by Chernyak and Zhitnitsky
of a "double-humped" wave function of the pion at a low scale, far
from the asymptotic form, was based on an extraction of the first
few moments from a standard QCD sum rule approach[5], which has
been criticized and revised in Refs.[6,7]. Subsequently, a number
of authors have proposed and studied the modified versions of
meson [7,8] and baryon wave functions [9,10]. Additional arguments
in favour of a form of the pion wave functions close to the
asymptotic one have come from the analysis of the transition form
factor $\gamma\gamma^{\star}\to\pi^0$ [12]. The measurements of
this form factor by the CLEO collaboration are consistent with a
near-asymptotic form of the wave function[15]. In [16], the
leading-twist wave function of the pion at a low normalization
point is calculated in the effective low-energy theory derived
from the instanton vacuum. These results for the pion wave
function at the low normalization point are close to the
asymptotic form and consistent with the CLEO measurements. The
authors have obtained a shape substantially different from the
Chernyak-Zhitnitsky one because they have chosen a significantly
smaller value of the second moment, and, more importantly, they
have taken  all the moments of the wave function into account.
Their results support the conclusions reached previously in
Refs.[6,7]. The QCD factorization theorems predict that the
hadron-hadron cross section can be obtained by the convolution of
parton distribution functions and a cross section of the
corresponding hard scattering subprocess. The parton distributions
are nonperturbative, process-independent quantities, which are
specific to any given hadron. The hard scattering cross sections
are independent of all long distance effects and can be found by
means of pQCD. In the framework of pQCD, the higher order
corrections to the hard scattering, and therefore to the
hadron-hadron process cross sections, have been calculated [17].
These corrections are large and change the leading order results
considerably. Other corrections to the hadron-hadron process cross
sections and its different characteristics come from the higher
twist (HT) terms.
 By taking these points into account, it
may be asserted that the analysis of the higher twist effects  on
the dependence of the meson wave function  in single pseudoscalar and vector
meson production at photon-photon collisions are significant in both
theoretical and experimental studies. Much effort has recently
been devoted to the study of exclusive processes involving large
transverse momenta within the context of perturbative quantum
chromodynamics (QCD). Here, as in other applications of
perturbative QCD, photon-induced reactions play an important role.
In [18] have been studied photon-photon anihilation into two
mesons at large center-of-mass angles $\theta_{c.m.}$. The
higher-twist contributions to high-$p_T$ inclusive meson
production in two-photon collisions, a single meson inclusive
photoproduction and jet photoproduction cross sections were
studied by various authors [19-21]. As experiments examining
high-$p_T$ particle production in two-photon collisions are
improved, it becomes important to reassess the various
contributions which arise in quantum chromodynamics. Predictins
for the higher-twist contributions, orginally obtained in Ref.22,
may now be refined using the exclusive-process QCD formalism
devoloped in [23]. Another important aspect of this study is the
choice of the meson model wave functions. In this respect, the
contribution of the high twist Feynman diagrams to a single meson
production cross section in photon-photon collisions has been
computed by using various meson wave functions. Also, the leading
and high twist contributions have been estimated and compared to
each other. Within this context, this paper is organized as
follows: in section \ref{ht}, we provide some formulae for the
calculation of the contribution of the high twist diagrams. In
section  \ref{lt}, we provide the formulae for the calculation of
the contribution of the leading twist diagrams and in section
\ref{results}, we present the numerical results for the cross
section and discuss the dependence of the cross section on the
meson wave functions. We state our conclusions in section
\ref{conc}.

\section{CONTRIBUTION OF THE HIGH TWIST DIAGRAMS}\label{ht}
The high twist Feynman diagrams, which describe the subprocess
$\gamma q \to M q $ contributes to $\gamma\gamma \to MX$ for the
meson production in the photon-photon collision are shown in
Fig.1(a). The amplitude for this subprocess can be found by means of
the Brodsky-Lepage formula [24]

\begin{equation}
M(\hat s,\hat
t)=\int_{0}^{1}{dx_1}\int_{0}^{1}dx_2\delta(1-x_1-x_2)\Phi_{M}(x_1,x_2,Q^2)T_{H}(\hat
s,\hat t;x_1,x_2).
\end{equation}

In Eq.(2.1), $T_H$ is  the sum of the graphs contributing to the
hard-scattering part of the subprocess. The hard-scattering part
for the subprocess under consideration is $\gamma q \to Mq$, in
which the observed meson is made directly. The hard-scattering amplitude $T_{H}(\hat s,\hat
t;x_1,x_2)$ depends on a process and can be obtained in the
framework of pQCD, whereas the wave function
$\Phi_{M}(x_1,x_2,Q^2)$ describes all the non-perturbative and
process-independent effects of hadronic binding. The hadron wave
function gives the amplitude for finding partons (quarks, gluons)
carrying the longitudinal fractional momenta
$\textbf{x}=(x_1,x_2,....x_n)$ and virtualness up to $Q^2$ within
the hadron and, in general, includes all Fock states with quantum
numbers of the hadron. But only the lowest Fock state
($q_1\bar{q}_{2}$-for mesons, $uud$-for proton, \emph{etc.})
contributes to the leading scaling behavior, other Fock state
contributions are suppressed by powers of $1/Q^2$. In our work, we
have restricted ourselves to considering the lowest Fock state for
a meson. Then $\textbf{x}= x_1,x_2$ and $x_1+x_2=1$. This approach
can be applied not only to the investigation of exclusive
processes[25], but also to the calculation of higher twist corrections
to some inclusive processes such as large-$p_T$ dilepton
production [26], two-jet+ meson production in the
electron-positron annihilation [27], ets. The
$q_{1}\overline{q}_{2}$ spin state used in computing $T_H$ may be
written in the form

\begin{equation}
\sum_{s_{1},s_{2}}
\frac{u_{s_1}({x}_{1}p_{M})\overline{v}_{s_{2}}({x}_{2}p_{M})}{\sqrt{x_1}
\sqrt{x_2}}\cdot N_{s_{1}s_{2}}^s=\left\{\begin{array}{ccc}
\frac{{\gamma}_{5}\hat {p}_{\pi}}{\sqrt{2}},\,\,\pi,\\\frac{\hat
{p}_{M}}{\sqrt{2}},\,\,\rho_L\,\,helicity \, 0,\\
\mp\frac{{\varepsilon}_{\mp}\hat {p}_{M}}{\sqrt{2}},\,\,
\rho{_T}\,\,helicity \pm1,\end{array}\right.
\end{equation}

where $\varepsilon_{\pm}=\mp(1/\sqrt{2})(0,1,\pm i,0)$ in a frame
with $(p_M)_{1,2}=0$ and the $N_{s_{1}s_{2}}^s$ project out a
state of spins $s$, and $p_{M}$ is the four-momentum of the final
meson. In our calculation, we have neglected the meson mass.
Turning to extracting the contributions of the high twist
subprocesses, there are many kinds of leading twist subprocesses
in $\gamma\gamma$ collisions as the background of the high twist subprocess
$\gamma q \to Mq$, such as $\gamma+\gamma \to
q+\overline{q}$. The contributions from these leading
twist subprocesses strongly depend on some phenomenological
factors, for example, quark and gluon distribution functions in
meson and fragmentation functions of various constituents
\emph{etc}. Most of these factors have not been well determined,
neither theoretically nor experimentally. Thus they cause very
large uncertainty in the computation of the cross section of
process $\gamma\gamma \to MX$. In general,
the magnitude of this uncertainty is much larger than the sum of
all the high twist contributions, so it is very difficult to
extract the high twist contributions.

The Mandelstam invariant variables for subprocesses
$\gamma q\to Mq$ are defined as
\begin{equation}
\hat s=(p_1+p_{\gamma})^2,\quad \hat t=(p_{\gamma}-p_{M})^2,\quad \hat
u=(p_1-p_{M})^2.
\end{equation}

In our calculation, we have also neglected the quark masses. We
have aimed to calculate the single meson production cross section
and to fix the differences due to the use of various meson wave
functions. We have used six different functions: the asymptotic
wave function ASY, the Chernyak-Zhitnitsky [2,5] and the wave
function in which two non-trivial Gegenbauer coefficients $a_2$
and $a_4$ have been extracted from the CLEO data on the
$\gamma\gamma^{\star} \to \pi^0$ transition form factor [28] and
Braun-Filyanov wave function [7] . In ref.[28], the authors have used
the QCD light-cone sum rules
approach and have included into their analysis the NLO
perturbative and twist-four corrections. They found that in the
model with two nonasymptotic terms, at the scale $\mu_0=2.4 GeV$. For
$\rho$-meson we used Ball-Braun wave function [29].
In order to proceed to numerical calculations we have to use the
explicit expressions for the $\rho$ meson wave functions. They are
defined by means of the following formulas: For the longitudinally
polarized $\rho$ meson, $\rho_{L}\equiv\rho_{h=0}$
\begin{equation}
<0\mid\overline{d}(z)\gamma_{\mu}u(-z)\mid\rho_{L}(p)>=f_{\rho}^{L}p_{\mu}\int_{-1}^{1}
d\xi e^{i\xi(zP)}\phi_{L}^{\rho}(\xi),
\end{equation}
for the transversely polarized $\rho$ meson, $\rho_{T}\equiv
\rho_{h=\pm1}$
\begin{equation}
<0\mid\overline{d}(z)\sigma_{\mu\nu}Iu(-z)\mid\rho_{T}(p)>=
f_{\rho}^{T}(\epsilon_{\mu}^{T}p_{\nu}-\epsilon_{\nu}^{T}
p_{\mu})\int_{-1}^{1}d\xi e^{i\xi(zP)}\phi_{T}^{\rho}(\xi),
\end{equation}
$f_{\rho}^{L}, f_{\rho}^{T}$ are the dimensional constants which
determine the values of the wave functions at the origin, and
\begin{equation}
I=exp\left[ig\int_{-z}^{z}d\sigma_{\mu}A^{\mu}(\sigma)\right]
\end{equation}
In Eqs.(2.4), (2.5) and (2.6), $\overline{d}(z)$, $u(z)$ and
$A_{\mu}(\sigma)$ are quark and gluon fields, $\epsilon_{\mu}^{T}$
is the polarization vector, and $|\rho_{L}(p)\rangle$,
$|\rho_{T}(p)\rangle$ are the longitudinally and transversely
polarized $\rho$ meson states with the momentum $p$.
$$
\Phi_{asy}(x)=\sqrt{3}f_{\pi}x(1-x),\quad
\Phi_{L(T)}^{asy}(x)=\sqrt{6}f_{\rho}^{L(T)}x(1-x) \\
$$
$$
\Phi_{CZ}(x,\mu_{0}^2)=5\Phi_{asy}(2x-1)^2,
\Phi_{L(T)}^{\rho}(x,\mu_{0}^2)=\Phi_{L(T)}^{asy}(x)\left[a+b(2x-1)^2\right],\quad\\
$$
$$
\Phi_{BF}(x,\mu_{0}^2)=\Phi_{asy}(x)[1+0.66(5(2x-1)^2-1)+0.4687(21(2x-1)^4-14(2x-1)^2+1)],
$$
\begin{equation}
\Phi_{CLEO}(x,\mu_{0}^2)=\Phi_{asy}(x)[1+0.285(5(2x-1)^2-1)-0.263(21(2x-1)^4-14(2x-1)^2+1)],
\end{equation}
where $f_{\pi}$=0.0923 GeV, $f_{\rho}^L$=0.141 GeV$,
f_{\rho}^{T}$=0.16 GeV is the pion and $\rho$ mesons  decay constants, $a=0.7$,
$b=1.5$ for both longitudinally and transversely polarized $\rho$
meson[29]. Here, we have denoted by $x\equiv x_1$, the longitudinal
fractional momentum carried by the quark within the meson. Then,
$x_2=1-x$ and $x_1-x_2=2x-1$. The pion and $\rho$ meson wave function is symmetric
under replacement $x_1-x_2\leftrightarrow x_2-x_1$. The values of
the pion wave function moments $<\xi^n>$ are defined as
\begin{equation}
<\xi^n>=\int_{-1}^{1}d\xi \xi^n\widetilde{\Phi}_{\pi}(\xi)
\end{equation}
Here, $\widetilde{\Phi}_{\pi}(\xi)$ is the model function without
$f_{\pi}$ and  $\xi=x_1-x_2$. The pion wave function moments have
been calculated by means of the QCD sum rules method by Chernyak
and Zhitnitsky at the normalization point $\mu_{0}=0.5GeV$. They
are equal to
\begin{equation}
<\xi^0>_{{\mu}_{0}}=1,\,\,\,<\xi^2>_{{\mu}_{0}}=0.44,\,\,<\xi^4>_{{\mu}_{0}}=0.27
\end{equation} The Chernyak-Zhitnitsky pion model wave function has
the following moments
\begin{equation}
<\xi^0>_{{\mu}_{0}}=1,\,\,\,<\xi^2>_{{\mu}_{0}}=0.43,\,\,<\xi^4>_{{\mu}_{0}}=0.24
\end{equation}
It is interesting to note that the corresponding moments of the
asymptotic wave function differ considerably from those in
Eqs.(2.9), (2.10)
\begin{equation}
<\xi^0>_{{\mu}_{0}}=1,\,\,\,<\xi^2>_{{\mu}_{0}}=0.20,\,\,<\xi^4>_{{\mu}_{0}}=0.086
\end{equation}
 This means that the realistic pion wave function is much wider than
the asymptotic one [5,30].
 The model functions can be written as
$$
\Phi_{asy}(x)=\sqrt{3}f_{\pi}x(1-x),
$$
$$
\Phi_{CZ}(x,\mu_{0}^2)=\Phi_{asy}(x)\left[C_{0}^{3/2}(2x-1)+\frac{2}{3}C_{2}^{3/2}(2x-1)\right],
$$
$$
\Phi_{L(T)}^{\rho}(x,\mu_{0}^2)=\Phi_{L(T)}^{asy}(x)\left[C_{0}^{3/2}(2x-1)+0.18(0.2)\frac{2}{3}C_{2}^{
3/2}(2x-1)\right],
$$
$$
\Phi_{BF}(x,\mu_{0}^2)=\Phi_{asy}(x)\left[C_{0}^{3/2}(2x-1)+0.44C_{2}^{3/2}(2x-1)+0.25C_{4}^{3/2}(2x-1)
\right]
$$
$$
\Phi_{CLEO}(x,\mu_{0}^2)=\Phi_{asy}(x)\left[C_{0}^{3/2}(2x-1)+0.19C_{2}^{3/2}(2x-1)-0.14C_{4}^{3/2}(2x-
1)\right],
$$
$$
C_{0}^{3/2}(2x-1)=1,\,\,C_{2}^{3/2}(2x-1)=\frac{3}{2}(5(2x-1)^2-1),
$$
\begin{equation}
C_{4}^{3/2}(2x-1)=\frac{15}{8}(21(2x-1)^4-14(2x-1)^2+1).
\end{equation}

It may be seen that the pion wave function extracted from the
experimental data depends on the methods used and their accuracy.
Although one may claim that the meson wave function is a
process-independent quantity, describing the internal structure of
the meson itself, the exploration of different exclusive processes
with the same meson leads to a variety of wave functions. This
means that the methods employed have shortcomings or do not
encompass all the mechanisms important for a given process. Such a
situation is pronounced in the case of the pion. It is known that
the meson wave function (distribution amplitude-DA) can be expanded
over the eigenfunctions of the one-loop Brodsky-Lepage equation,
\emph{i.e.}, in terms of the Gegenbauer polynomials
$\{C_{n}^{3/2}(2x-1)\},$
\begin{equation}
\Phi_{M}(x,Q^2)=\Phi_{asy}(x)\left[1+\sum_{n=2,4...}^{\infty}a_{n}(Q^2)C_{n}^{3/2}(2x-1)\right],
\end{equation}

The evolution of the wave function (DA) on the factorization scale
$Q^2$ is governed by the functions $a_n(Q^2)$,
\begin {equation}
a_n(Q^2)=a_n(\mu_{0}^2)\left[\frac{\alpha_{s}(Q^2)}{\alpha_{s}(\mu_{0}^2)}\right]^{\gamma_n/\beta_0},
\end{equation}
$$
\frac{\gamma_2}{\beta_{0}}=\frac{50}{81},\,\,\,\frac{\gamma_4}{\beta_{0}}=\frac{364}{405},\,\,
n_f=3.
$$
In Eq.(2.14), $\{\gamma_n\}$ are anomalous dimensions defined by
the expression,
\begin{equation}
\gamma_n=C_F\left[1-\frac{2}{(n+1)(n+2)}+4\sum_{j=2}^{n+1}
\frac{1}{j}\right].
\end{equation}

The constants $a_n(\mu_{0}^2)=a_{n}^0$ are input parameters that
form the shape of the wave functions and which can be extracted
from experimental data or obtained from the nonperturbative QCD
computations at the normalization point $\mu_{0}^2$. The QCD
coupling constant $\alpha_{s}(Q^2)$ at the two-loop approximation
is given by the expression

\begin{equation}
\alpha_{s}(Q^2)=\frac{4\pi}{\beta_0
ln(Q^2/\Lambda^2)}\left[1-\frac{2\beta_1}{\beta_{0}^2}\frac{lnln(Q^2/\Lambda^2)}{ln(Q^2/\Lambda^2)}\right].
\end{equation}
Here, $\Lambda$ is the QCD scale parameter, $\beta_0$ and
$\beta_1$ are the QCD beta function one- and two-loop
coefficients, respectively,
$$
\beta_0=11-\frac{2}{3}n_f,\,\,\,\beta_1=51-\frac{19}{3}n_f.
$$

For completennes we give the sum rules for the $\rho$- meson wave functions moments
$\langle\xi\rangle=\int d\xi\xi^{n}\phi(u,\mu)$ [29].
$$
(f_{\rho}^{\perp})^2(\mu)\langle\xi^{n}\rangle_{\perp}e^{{-m_{\rho}^2}/M^2}=
\frac{3}{2\pi^2}\int_{0}^{s_{0}}ds\int_{0}^{1}du
e^{-s/M^2}u\overline{u}(2u-1)^{n}\cdot
$$
$$
\left\{1+\frac{\alpha_{s}}{3\pi}(6-\frac{\pi^2}{3}+2ln\frac{s}{\mu^2}+lnu+ln\overline{u}+
ln^{2} \frac{u}{\overline{u}})\right\}+\frac{n-1}{n+1}
\frac{1}{12M^2}\langle\frac{\alpha_{s}}{\pi} G^2\rangle +
$$

\begin{equation}
\frac{64\pi}{81M^4}(n-1)\langle\sqrt{\alpha_{s}}\overline{q}q\rangle^2,
\end{equation}
$$
f_{\rho}^2\langle\xi\rangle_{\parallel}e^
{-m_{\rho^{2/M^2}}}=
\frac{3}{4\pi^{2}(n+1)(n+3)}(1+\frac{\alpha_{s}}{\pi}
A_{n}^{\prime})M^2(1-e^{-s_{0}/M^2})+
$$
\begin{equation}
\frac{1}{12M^2}\langle
\frac{\alpha_{s}}{\pi}G^2\rangle+\frac{16\pi}{81M^4}(4n-7)\langle\sqrt{\alpha_{s}}\overline{q}q\rangle^
2.
\end{equation}

Here the vacuum condensates are equal to [31]
$$
\langle \frac{\alpha_{s}}{\pi}G^2\rangle=(0.012\pm0.006)GeV^4\\
$$
$$
\langle\sqrt{\alpha_{s}}\overline{q}q\rangle^2=0.56(-0.25GeV)^6\\
$$

The higher-twist subprocess $\gamma q\to Mq$ contributes to
$\gamma\gamma \to MX$ through the diagram of Fig.1(a). We now incorporate the
higher-twist(HT) subprocess $\gamma q\to Mq$ into the full inclusive cross
section. In this subprocess $\gamma q\to Mq$, photon and the meson may be
viewed as an effective current striking the incoming quark line. With this in
mind, we write the complete cross section in formal analogy with
deep-inelastic scattering,

\begin{equation}
E\frac{d\sigma}{d^{3}p}(\gamma \gamma\to MX )=\frac
{3}{\pi}\sum_{q \overline{q}}\int_{0}^{1}dx \delta(\hat s+\hat
t+\hat u)\hat s G_{q/{\gamma}}(x,-\hat t)\frac{d\sigma}{d\hat
t}(\gamma q \to Mq)+ (t\leftrightarrow u),
\end{equation}

Here $G_{q/\gamma}$ is the per color distribution function for a
quark in a photon. The subprocess cross section for $\pi,\rho_{L}$
and $\rho_{T}$ production
\begin{equation}
\frac{d\sigma}{d\hat t}(\gamma q\to Mq)=\left\{\begin{array}{cc}
\frac{8\pi\alpha_{E}C_{F}}{9}[D(\hat s,\hat u)]^{2}
\frac{1}{\hat{s}^2(-\hat t)}\left[\frac{1}{\hat{s}^2}+
\frac{1}{\hat{u}^2}\right],\,\,\, M=\pi,\rho_{L},\\
\frac{8\pi\alpha_{E}C_{F}}{9} \left[D(\hat s,\hat
u)\right]^2\frac{8(-\hat t)}{\hat{s}^4 \hat{u}^2},M=\rho_{T},
\end{array} \right.
\end{equation}
 where
$$
 D(\hat s,\hat u)=\hat ue_{1}\alpha_{s}\left(\frac{\hat
 s}{2}\right)I_{M}\left(\frac{\hat s}{2}\right)+ \hat
se_{2}\alpha_{s}\left(\frac{-\hat
 u}{2}\right)I_{M}\left(\frac{-\hat u}{2}\right
 )
$$
where $Q_{1}^2=\hat
s/2,\,\,\,\,Q_{2}^2=-\hat u/2$,\,\, represents the momentum squared
carried by the hard gluon in Fig.1(a), $e_1(e_2)$ is the charge of
$q_1(\overline{q}_2)$ and $C_F=\frac{4}{3}$.

The $I_M$ factors reflect the exclusive form factor of the meson
and are discussed thoroughly in [20], as is the motivation the
arguments of $\alpha_{s}$ and $I_M$. Note that the relation
between $I_M$ and the meson form factor completely fixes the
normalization of the higher-twist subprocess. The full cross
section for $\pi$ and $\rho_{L}$ production is given by
$$
E\frac{d\sigma}{d^{3}p}(\gamma \gamma\to MX )=\frac{s}{s+u}
\sum_{q\overline{q}}G_{q/{\gamma}}(x,-\hat
t)\frac{8\pi\alpha_{E}C_{F}}{3}\frac{[D(\hat s,\hat u)]^2}{{\hat
s}^2(-\hat t)}\left[\frac{1}{{\hat s}^2}+\frac{1}{{\hat
u}^2}\right]+
$$
\begin{equation}
\frac{s}{s+t} \sum_{q\overline{q}}G_{q/{\gamma}}(x,-\hat
u)\frac{8\pi\alpha_{E}C_{F}}{3}\frac{[D(\hat s,\hat t)]^2}{{\hat
s}^2(-\hat u)}\left[\frac{1}{{\hat s}^2}+\frac{1}{{\hat
t}^2}\right] ,
\end{equation}
In (2.21), the subprocess invariants are
$$\hat s=xs,$$
$$\hat t=t,$$
\begin{equation}
\hat u=xu,
\end{equation}
$$t=-\frac{s}{2}(x_R-x_F)= -m_T \sqrt{s} e^{-y},$$
$$u=-\frac{s}{2}(x_R+x_F)= -m_T \sqrt {s} e^y,$$
with $x_R=(x_{F}^2+x_{T}^2)^{1/2}$. Here
$x_F=2(p_M)_{\parallel}/\sqrt s$ and $x_T=2(p_M)_{\perp}/\sqrt
s=2p_{T}/\sqrt s$ specify the longitudinal and transverse momentum
of the meson. In terms of these the rapidity of $M$ is given by
$$
y=\frac{1}{2}[(x_R+x_F)/(x_R-x_F)]
$$
where $m_T$ -- is the transverse mass of meson, which is given by
$$m_T=\sqrt{m^2+p_{T}^2}$$
As seen from (2.20) the subprocess cross section for longitudinal
$\rho_{L}$ production is very similiar to that for $\pi$ production,
but the transverse $\rho_{T}$ subprocess cross section  has a quite different form.
 We have extracted the following high twist subprocesses
contributing to the two covariant cross sections in Eq.(2.19)
\begin{equation}
\gamma q_{1}\to(q_1\overline{q}_2)q_1 \,\,\,\,\,,
\gamma\overline{q}_{2}\to(q_1\overline{q}_2)\overline{q}_2
\end{equation}

As seen from Eq.(2.21), at fixed $p_T$, the cross section falls very
slowly with $s$. Also, at fixed $s$, the cross section decreases
as $1/p_{T}^5$, multiplied by a slowly varying logarithmic
function which vanishes at the phase-spase boundary. Thus, the
$p_T$ spectrum is  fairly independent of $s$ expect near the
kinematic limit.

\section{CONTRIBUTION OF THE  LEADING TWIST DIAGRAMS}\label{lt}
Regarding the high twist corrections to the meson production cross
section, a comparison of our results with leading twist
contributions is crucial. The contribution from the leading-twist
subprocess $\gamma\gamma\to q\overline{q}$ is shown in Fig.1(b).
The corresponding  inclusive cross section for production of a
meson $M$ is given by
\begin{equation}
\left[\frac{d\sigma}{d^{3}p
}\right]_{\gamma\gamma \to
MX}=\frac{3}{\pi}\sum_{q,\overline{q}}\int_{0}^{1}\frac{dz}{z^2}
\delta(\hat{s}+\hat{t}+\hat{u})\hat{s}D_{q}^{M}(z,-\hat{t})\frac{d\sigma}
{d\hat{t}}(\gamma\gamma\to q\overline{q})
\end{equation}
where
$$
 \hat{s}=s,\,\,\hat{t}=\frac{t}{z}\,\,\,\hat{u}=\frac{u}{z}
$$
Here $s$, $t$, and $u$ refer to the overall $\gamma\gamma\to MX$
reaction. $D_{q}^{M}(z,-\hat t)$ represents
the quark fragmentation function into a meson containing a quark
of the same flavor. For $\pi^{+}$ production we assume
$D_{\pi^{+}/u}=D_{\pi^{+}/\overline{d}}$. In the leading twist subprocess, meson
is indirectly emitted from the quark with fractional momentum $z$. The
$\delta$ function may be expressed in terms of the parton
kinematic variables, and the $z$ integration may then be done. The final form for the
leading-twist contribution to the large-$p_{T}$ meson production
cross section in the process $\gamma\gamma\to MX$ is
$$
\Sigma_{M}^{LT}\equiv
E\frac{d\sigma}{d^{3}P}=\frac{3}{\pi}\sum_{q,\overline{q}}\int_{0}^{1}\frac{dz}{z^2}\delta
(\hat{s}+\hat{t}+\hat{u})\hat{s}D_{q}^{M}(z,-\hat{t})\frac{d\sigma}
{d\hat{t}}(\gamma\gamma\to q\overline{q})=
$$
\begin{equation}
\frac{3}{\pi}\sum_{q,\overline{q}}\int_{0}^{1}d\frac{1}{z}\delta(s+\frac{1}{z}
(t+u))\hat{s}D_{q}^{M}(z,-\hat{t})\frac{d\sigma}{d\hat{t}}(\gamma\gamma\to
q\overline{q})=
\frac{34}{27}\alpha_{E}^2\frac{1}{z}D_{q}^{M}(z)\frac{1}{{\hat
s}^2}\left[\frac{\hat t}{\hat u}+\frac{\hat u}{\hat t}\right]
\end{equation}
where
$$
 z=-\frac{t+u}{s}
$$

We should note that $D(z,-\hat t)/z$ behaves as $1/z^2$ as
$z\rightarrow0$. For the kinematic range considered in our
numerical calculations, $D(z,-\hat t)/z$ increases even more
rapidly. We obtain of the final cross section, Eq.(3.2), are as
follows: At fixed $p_T$, the cross section decreases with $s$
asymptotically as $1/s$. At fixed $s$, the $D(z,-\hat t)$ function causes the
cross section to decrease rapidly as $p_T$ increases towards the
phase-spase boundary $(z\rightarrow1)$. As $s$ increases, the phase-spase
boundary  moves to higher $p_T$, and the $p_T$
distribution broadens.

\section{NUMERICAL RESULTS AND DISCUSSION}\label{results}

In this section, the numerical results for higher twist effects on
the dependence of the chosen meson wave functions in the process
$\gamma \gamma \to MX$ are discussed. We have calculated the
dependence on the meson wave functions for the high twist
contribution to the  large-$p_T$ single pseudoscalar ${\pi^{+}}$ and
vector $\rho_{L}^{+}$, $\rho_{T}^{+}$ mesons production cross section
in the photon-photon collision . The ${\pi^{-}}$, ${\rho_{L}^{-}}$,
${\rho_{T}^{-}}$  cross sections are, of course, identical. In the
calculations, the asymptotic $\Phi_{asy}$, Chernyak-Zhitnitsky
$\Phi_{CZ}$, Braun-Filyanov wave function[7], and also, the pion
wave function, from which two non-trivial Gegenbauer coefficients
$a_2$ and $a_4$ have been extracted from the CLEO data on the
$\pi^{0}\gamma$ transition form factor have been used[20]. For
$\rho$-meson we used Ball-Braun wave function[29]. In the ref.[28],
authors have used the QCD light-cone sum rules approach and included
into their analysis the NLO perturbative and twist-four corrections.
For the high twist subprocess, we take $\gamma q \to Mq$ and we have
extracted the following two high twist subprocess $\gamma
q_{1}\to(q_1\overline{q}_2)q_1$, $\gamma\overline{q}_{2}\to(q_1\overline{q}_2)\overline{q}_2$
contributing to $\gamma\gamma\to MX$ cross sections. Inclusive meson
photoproduction represents a significant test case in which
higher-twist terms dominate those of leading twist in certain
kinematic domains. For the dominant leading twist subprocess for the
meson production, we take the photon-photon annihilation
$\gamma\gamma \to q\bar{q}$, in which the $M$ meson is indirectly
emitted from the quark. As an example for the quark distribution
function inside the photon  has been used [32]. The quark
fragmentation function has been taken from [33]. The other problems
dealth with are the choice of the QCD scale parameter $\Lambda$ and
the number of the active quark flavors $n_f$. The high twist
subprocesses probe the meson wave functions over a large range of
$Q^2$ squared momentum transfer, carried by the gluon. Therefore, we
take $Q_{1}^2=\hat s/2$, $Q_{2}^2=-\hat u/2$ which we have obtained
directly from the high twist subprocesses diagrams. The same $Q^2$
has been used as an argument of $\alpha_s(Q^2)$ in the calculation
of each diagram. The results of our numerical calculations are
plotted in Figs.2-7. Figs.2-3 show the dependence of the
differential cross sections of the high twist $\Sigma_{M}^{HT}$, and
ratio $R=\Sigma_{M}^{HT}$/$\Sigma_{M}^{LT}$ as a function of the
meson transverse momentum $p_T$ for six different meson wave
functions. As shown in Fig.2, the high twist differential cross
section is monotonically decreasing with an increase in the
transverse momentum of the meson. As seen from Fig.2, in all wave
functions of the mesons, the dependencies of the high twist cross
sections on the $p_T$ transverse momentum of the meson demonstrate
the same behavior. Also, as seen from Fig.2 the leading twist cross section is 2-4 order suppress the 
high
twist cross section in magnitude, on the dependence wave functions of
meson, respectively. On the other hand, the
higher twist corrections are very sensitive to the choice of the
meson wave function. We should note that the magnitude of the high
twist cross section in the pion wave function $\Phi_{CLEO}(x,Q^2)$
case is very close to the asymptotic wave function $\Phi_{asy}(x)$
case. In Fig.3, the ratio $R=\Sigma_{M}^{HT}$/$\Sigma_{M}^{LT}$ is
plotted at $y=0$ as a function of the meson transverse momentum
$p_T$ for the different meson wave functions. First of all, it is
seen that the values of $R$ for fixed $y$ and $\sqrt s$ depend on
the choice of the meson wave function. Also, the distinction between
$R(\Phi_{asy}(x))$ with $R(\Phi_{CLEO}(x,Q^2))$,
$R(\Phi_{CZ}(x,Q^2))$, $R(\Phi_{BF}(x,Q^2))$,
$R(\Phi_{\rho_{L}}(x,Q^2))$ and $R(\Phi_{\rho_{T}}(x,Q^2))$   have
been calculated. We have found that the distinction
$R(\Phi_{asy}(x))$ and $R(\Phi_{CLEO}(x,Q^2))$ is small, whereas a
distinction  between $R(\Phi_{asy}(x))$ with $R(\Phi_{CZ}(x,Q^2))$,
$R(\Phi_{BF}(x,Q^2))$,  $R(\Phi_{\rho_{L}}(x,Q^2))$ and
$R(\Phi_{\rho_{T}}(x,Q^2))$ is significant. For example, in the case
of $\sqrt s=89 GeV$, $y=0$, the distinction between
$R(\Phi_{asy}(x))$ with $R(\Phi_{i}(x,Q^2))$ $(i=CLEO, CZ, BF,
BB(L),BB(T))$ is shown in Table \ref{table1}. Thus, the distinction
between $R(\Phi_{asy}(x))$ and $R(\Phi_{CLEO}(x,Q^2))$ is maximum at
$p_T=14GeV/c$, but the distinction between $R(\Phi_{asy}(x))$ with
$R(\Phi_{CZ}(x,Q^2))$, $R(\Phi_{BF}(x,Q^2))$,
$R(\Phi_{\rho_{L}}(x,Q^2))$  $R(\Phi_{\rho_{T}}(x,Q^2))$ is maximum
at $p_T=2 GeV/c$ and decreases with an increase in $p_T$. Such a
behavior  of $R$ may  be explained by reducing all moments of the
meson model wave functions to those of $\Phi_{asy}(x)$ for high
$Q^2$. In Fig.4, the ratio $R=\Sigma_{M}^{HT}$/$\Sigma_{M}^{LT}$ is
plotted at $p_T=7GeV/c$ as a function of the  rapidity $y$ of the
meson for the different meson wave functions. As we are  now in the
high energy region, the change of the rapidity to determine these
relations is given by $-ln(\sqrt s/p_T)\leq y \leq ln(\sqrt s/p_T)$.
At $\sqrt s=89 GeV$ and $p_T=7GeV/c$, the meson rapidity lies in the
region $-2.543\leq y\leq2.543$. First of all, it is seen that the
values of $R$ for fixed $p_{T}$ and $\sqrt s$ depend on the choice
of the meson wave function. As shown in Fig.4 in
all wave functions of the mesons, the dependencies of the ratio
$R=\Sigma_{M}^{HT}$/$\Sigma_{M}^{LT}$ of the rapidity $y$ of the meson has a
minimum approximately at one point $y=1.75$. After this point, the ratio
increases with increasing $y$. Also, the distinction between
$R(\Phi_{asy}(x))$ with $R(\Phi_{CLEO}(x,Q^2))$,
$R(\Phi_{CZ}(x,Q^2))$, $R(\Phi_{BF}(x,Q^2))$,
$R(\Phi_{\rho_{L}}(x,Q^2))$ and $R(\Phi_{\rho_{T}}(x,Q^2))$   have
been calculated. We have found that the distinction
$R(\Phi_{asy}(x))$ and $R(\Phi_{CLEO}(x,Q^2))$ is small, whereas a
distinction  between $R(\Phi_{asy}(x))$ with $R(\Phi_{CZ}(x,Q^2))$,
$R(\Phi_{BF}(x,Q^2))$, $R(\Phi_{\rho_{L}}(x,Q^2))$ and
$R(\Phi_{\rho_{T}}(x,Q^2))$ is significant. For example, in the case
of $\sqrt s=89 GeV$, $p_T=7GeV/c$, the distinction between
$R(\Phi_{asy}(x))$ with $R(\Phi_{i}(x,Q^2))$ $(i=CLEO, CZ, BF,
BB(L),BB(T))$ is presented in Table \ref{table2}.
Thus, the distinction between $R(\Phi_{asy}(x))$ and
$R(\Phi_{CLEO}(x,Q^2))$ is maximum at $y= -2.25$, but the
distinction between $R(\Phi_{asy}(x))$ with $R(\Phi_{CZ}(x,Q^2))$,
$R(\Phi_{BF}(x,Q^2))$, $R(\Phi_{\rho_{L}}(x,Q^2))$ and $R(\Phi_{\rho_{T}}(x,Q^2))$  is maximum at
$y=1.75$ and decreases with an increase in $y$. Such a behavior  of
$R$ may  be explained by reducing all moments of the pion model wave
functions to those of $\Phi_{asy}(x)$ for high $Q^2$. We have also
carried out comparative calculations in the center-of-mass energy
$\sqrt s=209 GeV$. Figs.5-6 show the dependence of the differential
cross sections of the high twist $\Sigma_{M}^{HT}$, and ratio
$R=\Sigma_{M}^{HT}$/$\Sigma_{M}^{LT}$ as a function of the meson
transverse momentum $p_T$ for six different meson wave functions. As
shown in Fig.5, the high twist differential cross section is
monotonically decreasing with an increase in the transverse momentum
of the meson. As seen from Fig.5, in all wave functions of the
mesons, the dependencies of the high twist cross sections on the
$p_T$ transverse momentum of the meson demonstrate the same
behavior. Also, as seen from Fig.5 the leading twist cross section is 4-5 order suppress the high
twist cross section in magnitude, on the dependence wave functions of
meson, respectively. Also here, as in Fig.2 the higher
twist corrections are very sensitive to the choice of the meson wave
function. We should note that the magnitude of the high twist cross
section in the pion wave function $\Phi_{CLEO}(x,Q^2)$ case is very
close to the asymptotic wave function $\Phi_{asy}(x)$ case. In
Fig.6, the ratio $R=\Sigma_{M}^{HT}$/$\Sigma_{M}^{LT}$ is plotted at
$y=0$ as a function of the meson transverse momentum $p_T$ for the
different meson wave functions. First of all, it is seen that the
values of $R$ for fixed $y$ and $\sqrt s$ depend on the choice of
the meson wave function. Also, the distinction between
$R(\Phi_{asy}(x))$ with $R(\Phi_{CLEO}(x,Q^2))$,
$R(\Phi_{CZ}(x,Q^2))$, $R(\Phi_{BF}(x,Q^2))$,
$R(\Phi_{\rho_{L}}(x,Q^2))$ and $R(\Phi_{\rho_{T}}(x,Q^2))$   have
been calculated. We have found that the distinction
$R(\Phi_{asy}(x))$ and $R(\Phi_{CLEO}(x,Q^2))$ is small, whereas a
distinction  between $R(\Phi_{asy}(x))$ with $R(\Phi_{CZ}(x,Q^2))$,
$R(\Phi_{BF}(x,Q^2))$, $R(\Phi_{\rho_{L}}(x,Q^2))$ and
$R(\Phi_{\rho_{T}}(x,Q^2))$ is significant. For example, in the case
of $\sqrt s=209 GeV$, $y=0$, the distinction between
$R(\Phi_{asy}(x))$ with $R(\Phi_{i}(x,Q^2))$ $(i=CLEO, CZ, BF,
BB(L),BB(T))$ is shown in Table \ref{table3}. Thus, the distinction
between $R(\Phi_{asy}(x))$ with
$R(\Phi_{CLEO}(x,Q^2))$, $R(\Phi_{CZ}(x,Q^2))$ and
$R(\Phi_{BF}(x,Q^2))$ is maximum at $p_T=20GeV/c$, but the
distinction between $R(\Phi_{asy(LT)}(x))$ with
$R(\Phi_{\rho_{L}}(x,Q^2))$ and  $R(\Phi_{\rho_{T}}(x,Q^2))$ is
maximum at $p_T=100 GeV/c$ and decreases with an increase in $p_T$.
In Fig.7, the ratio $R=\Sigma_{M}^{HT}$/$\Sigma_{M}^{LT}$ is plotted
at $p_T=17GeV/c$ as a function of the  rapidity $y$ of the meson for
the different meson wave functions. At $\sqrt s=209 GeV$ and
$p_T=17GeV/c$, the meson rapidity lies in the region $-2.509\leq
y\leq2.509$. First of all, it is seen that the values of $R$ for
fixed $p_{T}$ and $\sqrt s$ depend on the choice of the meson wave
function. Also, the distinction between $R(\Phi_{asy}(x))$ with
$R(\Phi_{CLEO}(x,Q^2))$, $R(\Phi_{CZ}(x,Q^2))$,
$R(\Phi_{BF}(x,Q^2))$, $R(\Phi_{\rho_{L}}(x,Q^2))$ and
$R(\Phi_{\rho_{T}}(x,Q^2))$   have been calculated. We have found
that the distinction $R(\Phi_{asy}(x))$ and $R(\Phi_{CLEO}(x,Q^2))$
is small, whereas a distinction  between $R(\Phi_{asy}(x))$ with
$R(\Phi_{CZ}(x,Q^2))$, $R(\Phi_{BF}(x,Q^2))$,
$R(\Phi_{\rho_{L}}(x,Q^2))$ and $R(\Phi_{\rho_{T}}(x,Q^2))$ is
significant. For example, in the case of $\sqrt s=209 GeV$,
$p_T=17GeV/c$, the distinction between $R(\Phi_{asy}(x))$ with
$R(\Phi_{i}(x,Q^2))$ $(i=CLEO, CZ, BF, BB(L),BB(T))$ is shown in
Table \ref{table4}.
Thus, the distinction between $R(\Phi_{asy}(x))$ and
$R(\Phi_{CLEO}(x,Q^2))$ is maximum at $y=-2.25$, but the
distinction between $R(\Phi_{asy}(x))$ with $R(\Phi_{CZ}(x,Q^2))$,
$R(\Phi_{BF}(x,Q^2))$, $R(\Phi_{\rho_{L}}(x,Q^2))$ and  $R(\Phi_{\rho_{T}}(x,Q^2))$ is maximum at
$y=1.75 $ and decreases with an increase in $y$. Also, as seen from Fig.7 in
all wave functions of the mesons, the dependencies of the ratio
$R=\Sigma_{M}^{HT}$/$\Sigma_{M}^{LT}$ of the rapidity $y$ of the meson has a
minimum approximately at one point $y=1.75$. After this point, the ratio
increases with increasing $y$.
As seen from calculations with increasing center-of-mass  energy
from $\sqrt s=89GeV$ to $\sqrt s=209GeV$, the distinction between
$R$ decreases for all meson wave functions.

\section{Concluding Remarks}\label{conc}

In this work, we have calculated the higher twist contribution to
the large-$p_T$ meson production cross section to show  the
dependence on the chosen meson wave functions in the process $\gamma\gamma
\to MX$. In our calculations, we have
used the asymptotic $\Phi_{asy}$, Chernyak-Zhitnitsky $\Phi_{CZ}$,
Braun-Filyanov $\Phi_{BF}$ wave functions and
also, the pion wave function, in which the coefficients $a_{2}$
and $a_{4}$ have been extracted from the CLEO data on the
$\pi^{0}\gamma$ transition form factor used. For $\rho$-meson we used
Ball-Braun wave function.  For the high twist
subprocess, we have taken $\gamma q \to Mq$. We have extracted the following two
high twist subprocesses $\gamma q_{1}\to(q_1\overline{q}_2)q_1$, 
$\gamma\overline{q}_{2}\to(q_1\overline{q}_2)\overline{q}_2$
contributing to $\gamma\gamma\to MX$ cross
sections. As the dominant leading twist subprocess for the
meson production, we have taken the photon-photon annihilation
$\gamma\gamma \to q\bar{q}$, where the $M$ meson is indirectly
emitted from the quark. The results of our numerical calculations
have been plotted in Figs.2-7. As shown in Figs.2,5 the high twist differential cross section
monotonically decrease when the transverse momentum of the meson
increases. As seen from Figs.2,5 in all wave
functions of mesons, the dependencies of the high twist cross
sections on the $p_T$ transverse momentum of the meson demonstrate
the same behavior. But, the higher twist corrections are very
sensitive to the choice of the meson wave function. It should be
noted that the magnitude of the high twist cross section for the
pion wave function $\Phi_{CLEO}(x,Q^2)$ is very close to the
asymptotic wave function $\Phi_{asy}(x)$.

In Figs.3 and 6, the ratio
$R=\Sigma_{M}^{HT}$/$\Sigma_{M}^{LT}$ has been plotted
at $y=0$ as a function of the meson transverse momentum, $p_T$, for
the different meson wave functions. It may be observed that the
values of $R$ for fixed $y$ and $\sqrt s$ depend on the choice of
meson wave function. Within this context, we have also calculated
the distinction between $R(\Phi_{asy}(x))$ and
$R(\Phi_{CLEO}(x,Q^2))$, $R(\Phi_{CZ}(x,Q^2))$,
$R(\Phi_{BF}(x,Q^2))$, $R(\Phi_{\rho_{L}}(x,Q^2))$, $R(\Phi_{\rho_{T}}(x,Q^2))$ . We have
ultimately found that the difference between $R(\Phi_{asy}(x))$
and $R(\Phi_{CLEO}(x,Q^2))$ is small, whereas a distinction
between $R(\Phi_{asy}(x))$ with $R(\Phi_{CZ}(x,Q^2))$,
$R(\Phi_{BF}(x,Q^2))$, $R(\Phi_{\rho_{L}}(x,Q^2))$ and $R(\Phi_{\rho_{T}}(x,Q^2))$ is significant.
In Figs.4 and 7, the ratio
$R=\Sigma_{M}^{HT}$/$\Sigma_{M}^{LT}$ has been plotted
at $p_T=7,17GeV/c$ as a function of the rapidity of the meson  for
the different meson wave functions. It may be observed that the
values of $R$ for fixed $p_T$ and $\sqrt s$ depend on the choice of
meson wave function. Within this context, we have also calculated
the distinction between $R(\Phi_{asy}(x))$ and
$R(\Phi_{CLEO}(x,Q^2))$, $R(\Phi_{CZ}(x,Q^2))$,
$R(\Phi_{BF}(x,Q^2))$, $R(\Phi_{\rho_{L}}(x,Q^2))$, $R(\Phi_{\rho_{T}}(x,Q^2))$ . We have
ultimately found that the difference between $R(\Phi_{asy}(x))$
and $R(\Phi_{CLEO}(x,Q^2))$ is small, whereas a distinction
between $R(\Phi_{asy}(x))$ with $R(\Phi_{CZ}(x,Q^2))$,
$R(\Phi_{BF}(x,Q^2))$, $R(\Phi_{\rho_{L}}(x,Q^2))$ and $R(\Phi_{\rho_{T}}(x,Q^2))$ is significant.
Our investigation enables us to conclude that the high twist meson
production cross section in the photon-photon collisions depends
on the form of the meson model wave functions and may be used for
their study. Further investigations are needed in order to clarify
the role of high twist effects  in QCD.

\section*{Acknowledgments}
One of author, A. I.~Ahmadov would like to thank Prof. Hans Peter Nilles and
also other members of the Physikalisches Institute  for appreciates
hospitality extended to him in Bonn, where this work has been carried out. The
financial support by DAAD is gratefully acknowledged. A. I.~Ahmadov is also  grateful to
NATO Reintegration Grant-980779

\newpage

\begin{table}[h]
\begin{center}
\begin{tabular}{|c|c|c|c|c|c|c} \hline
$p_{T},GeV/c$ & $\frac{R(\Phi_{CLEO}(x,Q^2))}{R(\Phi_{asy}(x))}$ & 
$\frac{R(\Phi_{CZ}(x,Q^2))}{R(\Phi_{asy}(x))}$ &
$\frac{R(\Phi_{BF}(x,Q^2))}{R(\Phi_{asy}(x))}$ &
$\frac{R(\Phi_{BB(L)}(x,Q^2))}{R(\Phi_{asy}(x))}$ &
$\frac{R(\Phi_{BB(T)}(x,Q^2))}{R(\Phi_{asy}(x))}$ \\ \hline
  2 & 0.232  & 9.054 & 23.999  &3.490 & 4.048 \\ \hline
  6 & 1.492  & 4.216 & 5.872 &3.950 & 2.861  \\ \hline
  14& 1.773  & 1.747 & 1.216 & 1.394 & 1.537 \\ \hline
  22 & 0.9499  & 0.7607 & 0.68177  &0.85668 & 0.80928 \\ \hline
  28 & 0.5568  & 0.5023 & 0.7200 &0.68469 & 0.5859  \\ \hline
  36& 0.9874  & 0.79656 & 0.6965 &0.8766 & 0.8341 \\ \hline

\end{tabular}
\end{center}
\caption{The distinction between $R(\Phi_{asy}(x))$ with
$R(\Phi_{i}(x,Q^{2}))$  (i=CLEO, CZ, BF, BB(L), BB(T) ) at c.m.
 energy $\sqrt s=89GeV$.} \label{table1}
\end{table}

\begin{table}[h]
\begin{center}
\begin{tabular}{|c|c|c|c|c|c|c}\hline
$y$ & $\frac{R(\Phi_{CLEO}(x,Q^2))}{R(\Phi_{asy}(x))}$ &$\frac{R(\Phi_{CZ}(x,Q^2))}{R(\Phi_{asy}(x))}$ 
&
$\frac{R(\Phi_{BF}(x,Q^2))}{R(\Phi_{asy}(x))}$ &
$\frac{R(\Phi_{BB(L)}(x,Q^2))}{R(\Phi_{asy}(x))}$ &
$\frac{R(\Phi_{BB(T)}(x,Q^2))}{R(\Phi_{asy}(x))}$ \\ \hline
  -2.25 & 1.545 & 1.22 & 0.745  &1.113 & 1.146 \\ \hline
  -1.5 & 0.769  & 0.594 & 0.615 &0.773 & 0.714  \\ \hline
  0.25& 1.428  & 4.183 & 5.97 & 2.399 & 2.893 \\ \hline
  1.0 & 0.916  & 5.169 & 9.693  & 2.773 & 3.276 \\ \hline
  1.75 & 0.559  & 6.936 & 15.798 & 3.186 & 5.549  \\ \hline
  2.25& 0.892  & 4.922 & 9.281 & 2.744 & 3.529 \\ \hline

\end{tabular}
\end{center}
\caption{The distinction between $R(\Phi_{asy}(x))$ with
$R(\Phi_{i}(x,Q^{2}))$ (i=CLEO, CZ, BF, BB(L), BB(T)) at c.m.
energy $\sqrt s=89GeV$.} \label{table2}
\end{table}

\begin{table}[h]
\begin{center}
\begin{tabular}{|c|c|c|c|c|c|c}\hline
$p_{T},GeV/c$ & $\frac{R(\Phi_{CLEO}(x,Q^2))}{R(\Phi_{asy}(x))}$ & 
$\frac{R(\Phi_{CZ}(x,Q^2))}{R(\Phi_{asy}(x))}$ &
$\frac{R(\Phi_{BF}(x,Q^2))}{R(\Phi_{asy}(x))}$ &
$\frac{R(\Phi_{BB(L)}(x,Q^2))}{R(\Phi_{asy}(x))}$ &
$\frac{R(\Phi_{BB(T)}(x,Q^2))}{R(\Phi_{asy}(x))}$ \\ \hline
20 & 1.686  & 2.760 & 2.865  &1.914 & 2.298 \\ \hline
  35 & 1.582  & 1.507 & 1.088 &1.294 & 1.413  \\ \hline
  50& 1.014  & 0.835 & 0.7266 & 0.8962 & 0.8577 \\ \hline
  65 & 0.614  & 0.5534 & 0.7367  & 0.70467 & 0.6017 \\ \hline
  85 & 1.011  & 0.839 & 0.736 & 0.0.8977 & 0.8586  \\ \hline
  100& 1.574  & 3.368 & 4.118 & 2.170 & 2.637 \\ \hline

\end{tabular}
\end{center}
\caption{The distinction between $R(\Phi_{asy}(x))$ with
$R(\Phi_{i}(x,Q^{2}))$  (i=CLEO, CZ, BF, BB(L), BB(T) ) at c.m.
energy $\sqrt s=209 GeV$.}\label{table3}
\end{table}

\begin{table}[h]
\begin{center}
\begin{tabular}{|c|c|c|c|c|c|c}\hline
$y$ & $\frac{R(\Phi_{CLEO}(x,Q^2))}{R(\Phi_{asy}(x))}$ & $\frac{R(\Phi_{CZ}(x,Q^2))}{R(\Phi_{asy}(x))}$ 
&
$\frac{R(\Phi_{BF}(x,Q^2))}{R(\Phi_{asy}(x))}$ &
$\frac{R(\Phi_{BB(L)}(x,Q^2))}{R(\Phi_{asy}(x))}$ &
$\frac{R(\Phi_{BB(T)}(x,Q^2))}{R(\Phi_{asy}(x))}$ \\ \hline
  -2.25 & 1.571 & 1.383 & 0.933  &1.212 & 1.289 \\ \hline
  -1.5 & 0.7519  & 0.616 & 0.673 & 0.7689 & 0.6978  \\ \hline
  0.25& 1.435  & 3.484 & 4.626 & 2.236 & 2.754 \\ \hline
  1.0 & 1.039  & 4.255 & 7.252  & 2.575 & 3.149 \\ \hline
  1.75 & 0.7246  & 5.887 & 12.269 & 2.999 & 6.338  \\ \hline
  2.25& 1.056  & 4.003 & 6.684 & 2.520 & 3.295 \\ \hline

\end{tabular}
\end{center}
\caption{The distinction between $R(\Phi_{asy}(x))$ with
$R(\Phi_{i}(x,Q^{2}))$ (i=CLEO, CZ, BF, BB(L), BB(T)) at c.m.
energy $\sqrt s=209GeV$.} \label{table4}
\end{table}

\newpage

\begin{figure}[htpb]
\epsfxsize 8cm \centerline{\epsfbox{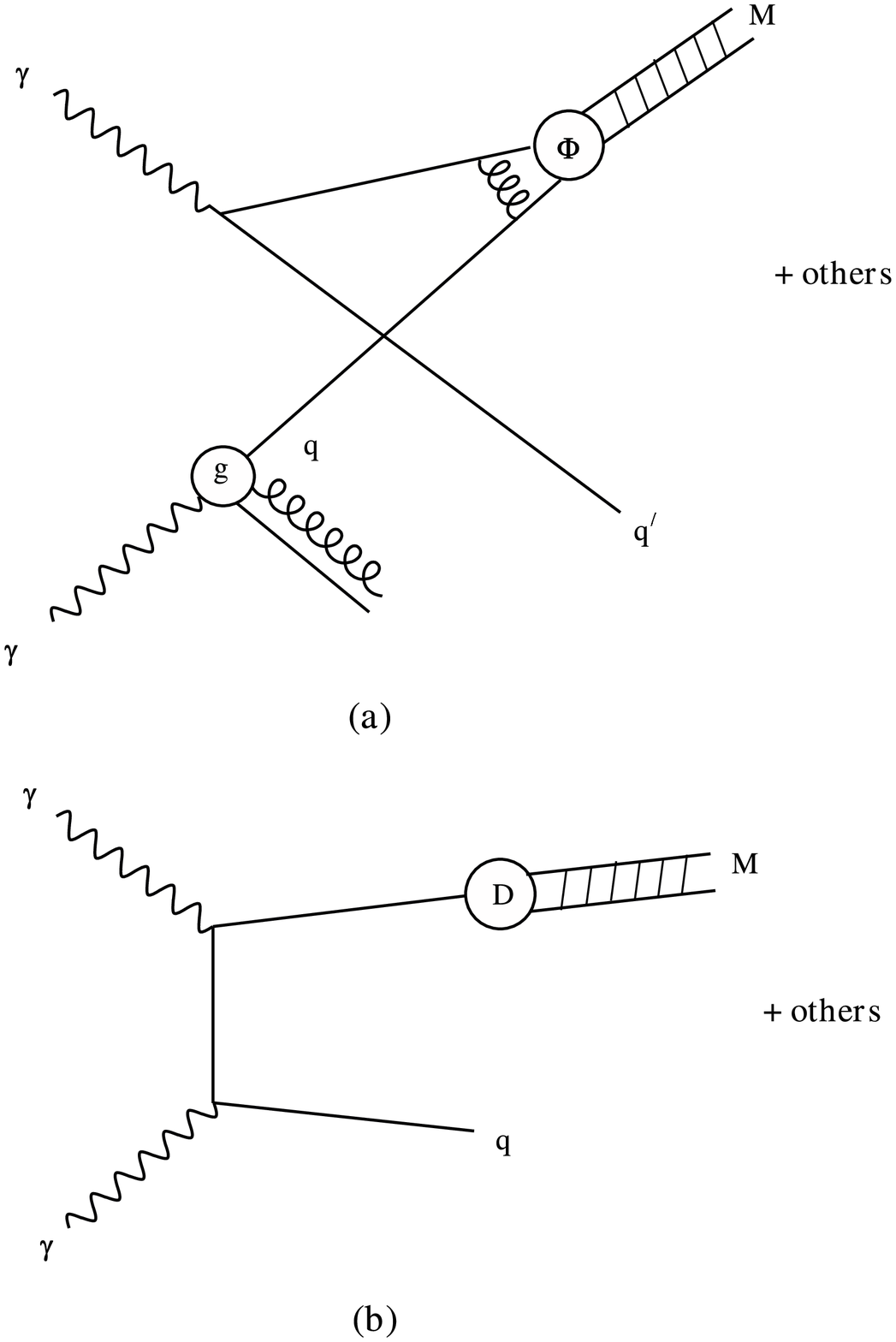}} \vskip -0.02cm
\caption{(a): The higher-twist contribution to $\gamma\gamma\to
MX$;\,\,\,(b): The leading-twist contribution to  $\gamma\gamma\to
MX$} \label{Fig1}
\end{figure}

\newpage

\begin{figure}[htb]
\vskip-2.2cm \epsfxsize 20cm \centerline{\epsfbox{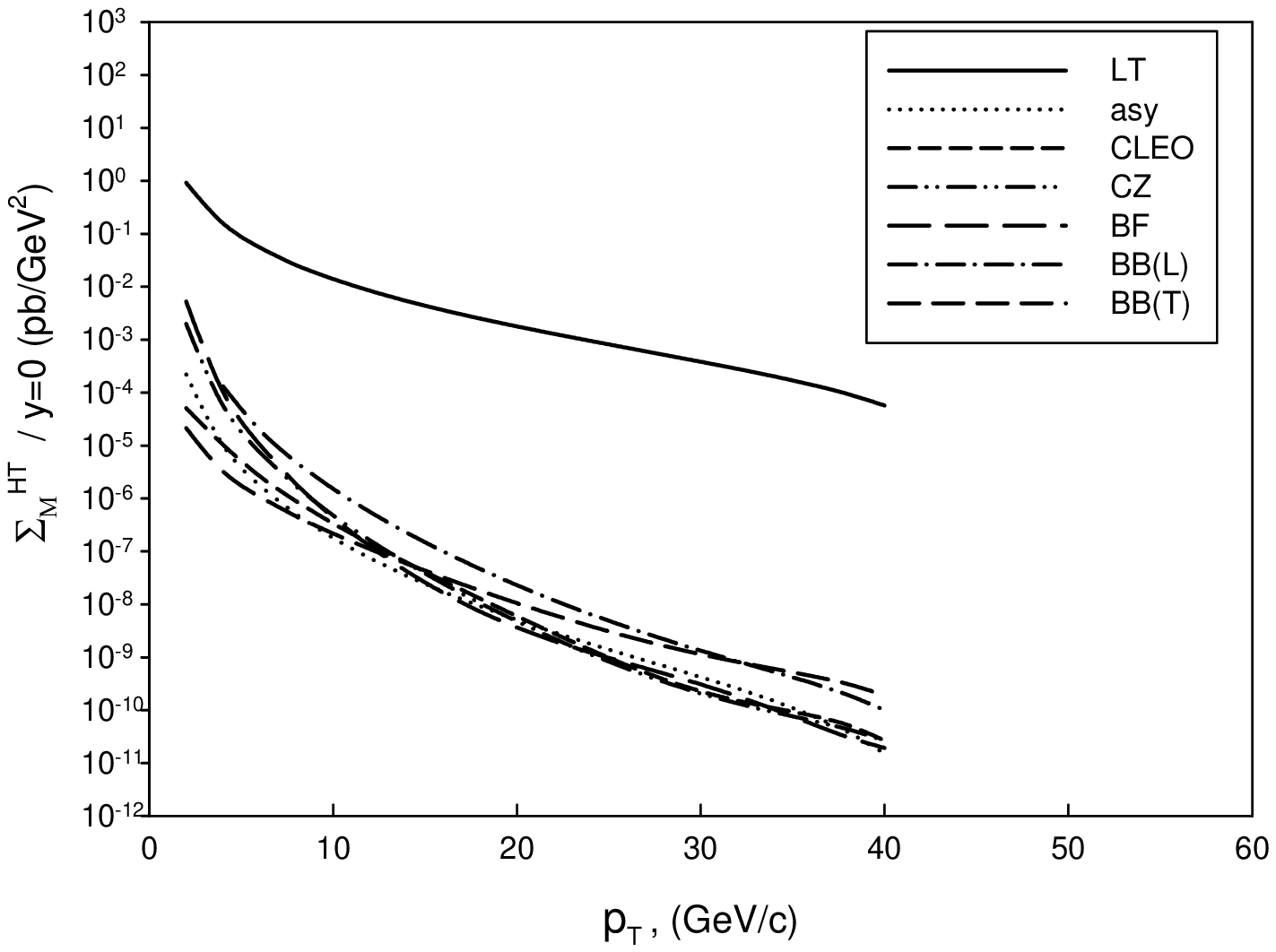}}
\vskip-18cm \caption{High twist  meson $M$ production cross
sections as a function of the $p_T$ transverse momentum of the
meson at the c.m.energy $\sqrt s=89 \,\,GeV$.} \label{Fig2}
\end{figure}

\begin{figure}[htb]
\vskip-1.8cm \epsfxsize 18cm \centerline{\epsfbox{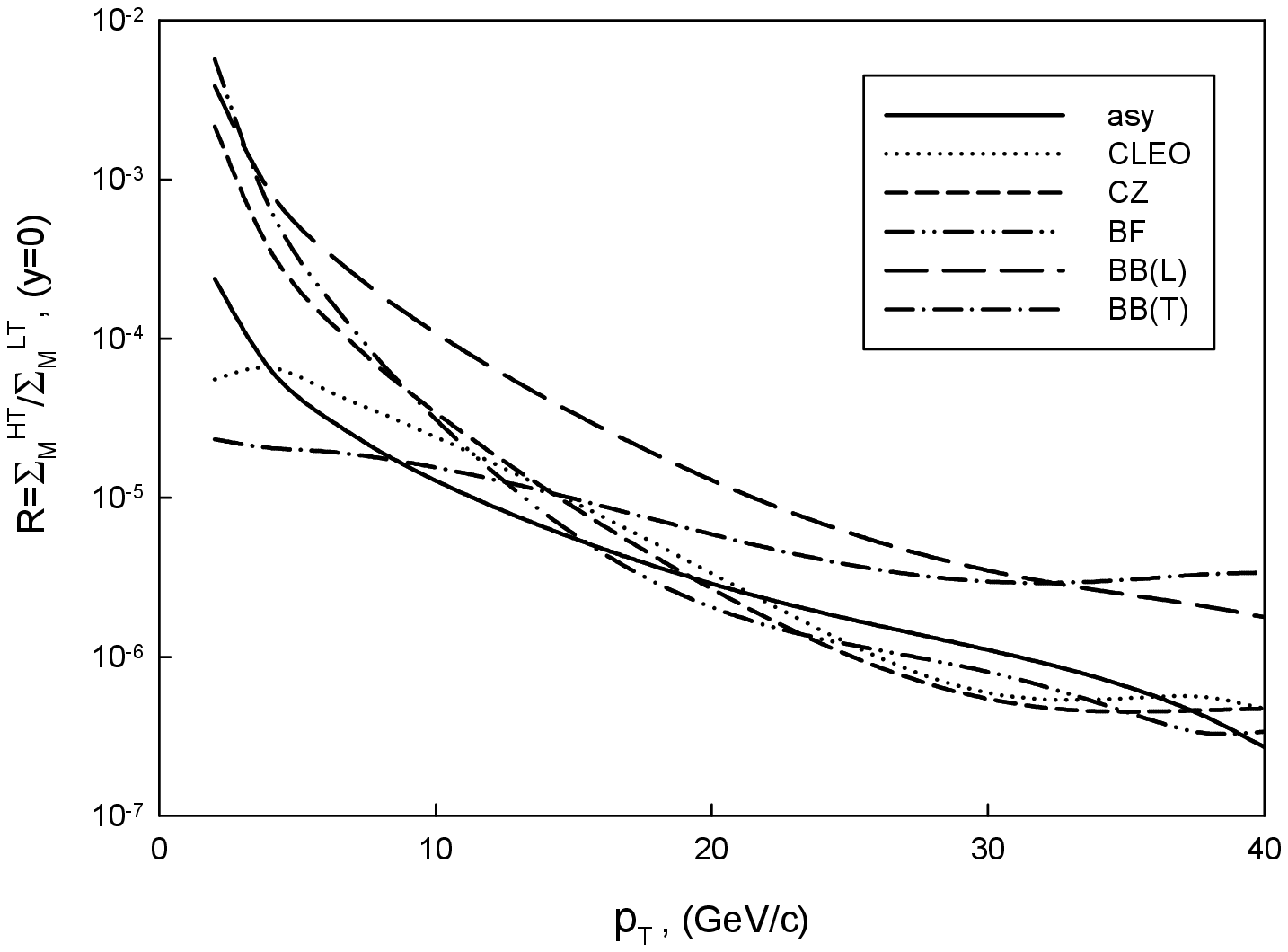}}
\vskip-16.5cm \caption{Ratio
$R=\Sigma_{M}^{HT}$/$\Sigma_{M}^{LT}$, where the leading and the
high twist contributions are calculated for the meson rapidity
$y=0$ at the c.m. energy $\sqrt s=89 \,\,GeV$, as a function of
the meson transverse momentum, $p_T$.} \label{Fig3}
\end{figure}

\newpage

\begin{figure}[htb]
\vskip-2.20cm\epsfxsize 19.5cm \centerline{\epsfbox{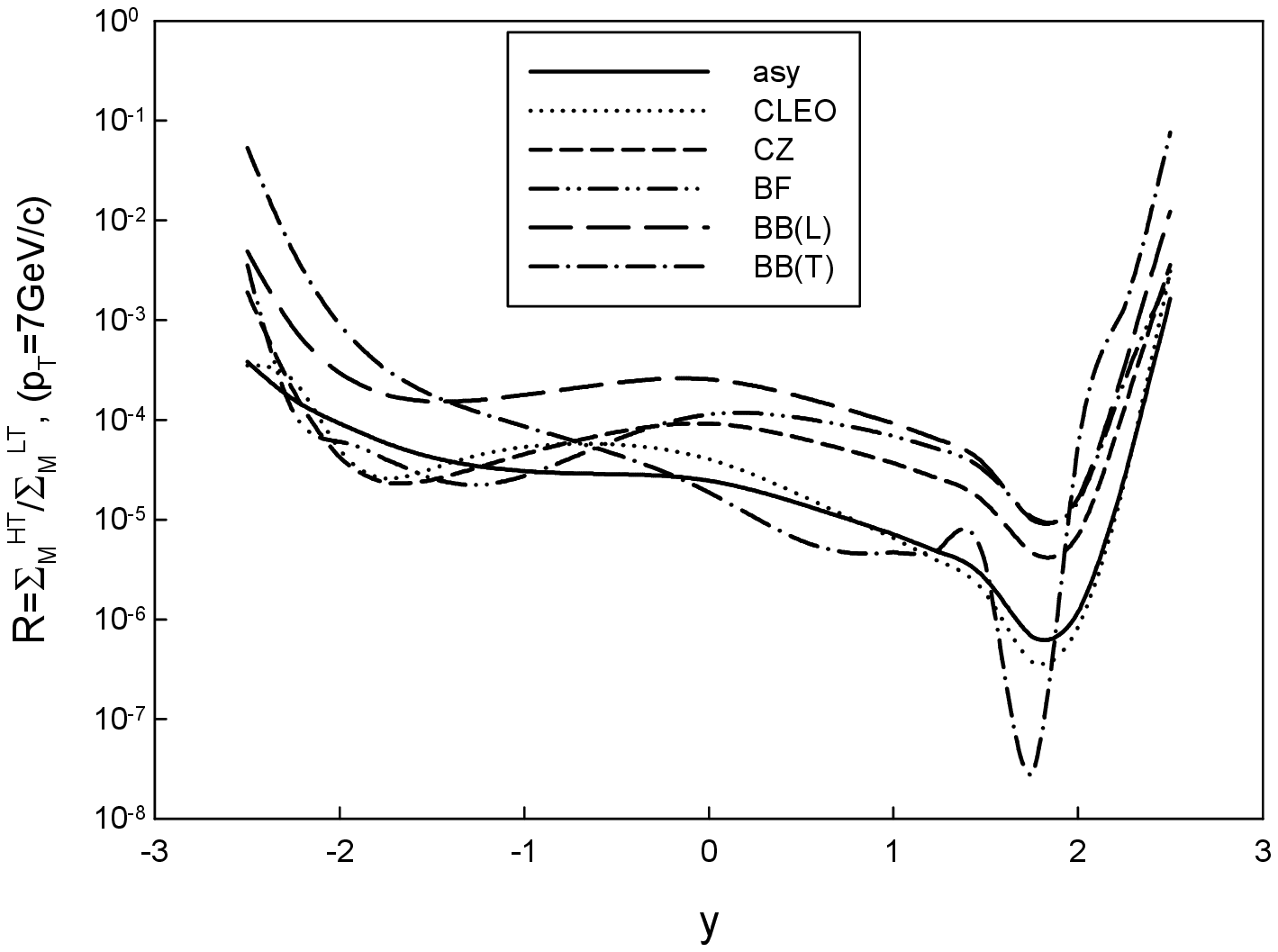}}
\vskip-17.7cm \caption{Ratio
$R=\Sigma_{M}^{HT}$/$\Sigma_{M}^{LT}$, where the leading and the
high twist contributions are calculated for the meson transverse
momentum $p_T=7 GeV/c$ at the c.m. energy $\sqrt s=89 \,\,GeV$, as
a function of the $y$ rapidity of the meson.} \label{Fig4}
\end{figure}

\begin{figure}[htb]
\vskip-2.1cm\epsfxsize 19.2cm
\centerline{\epsfbox{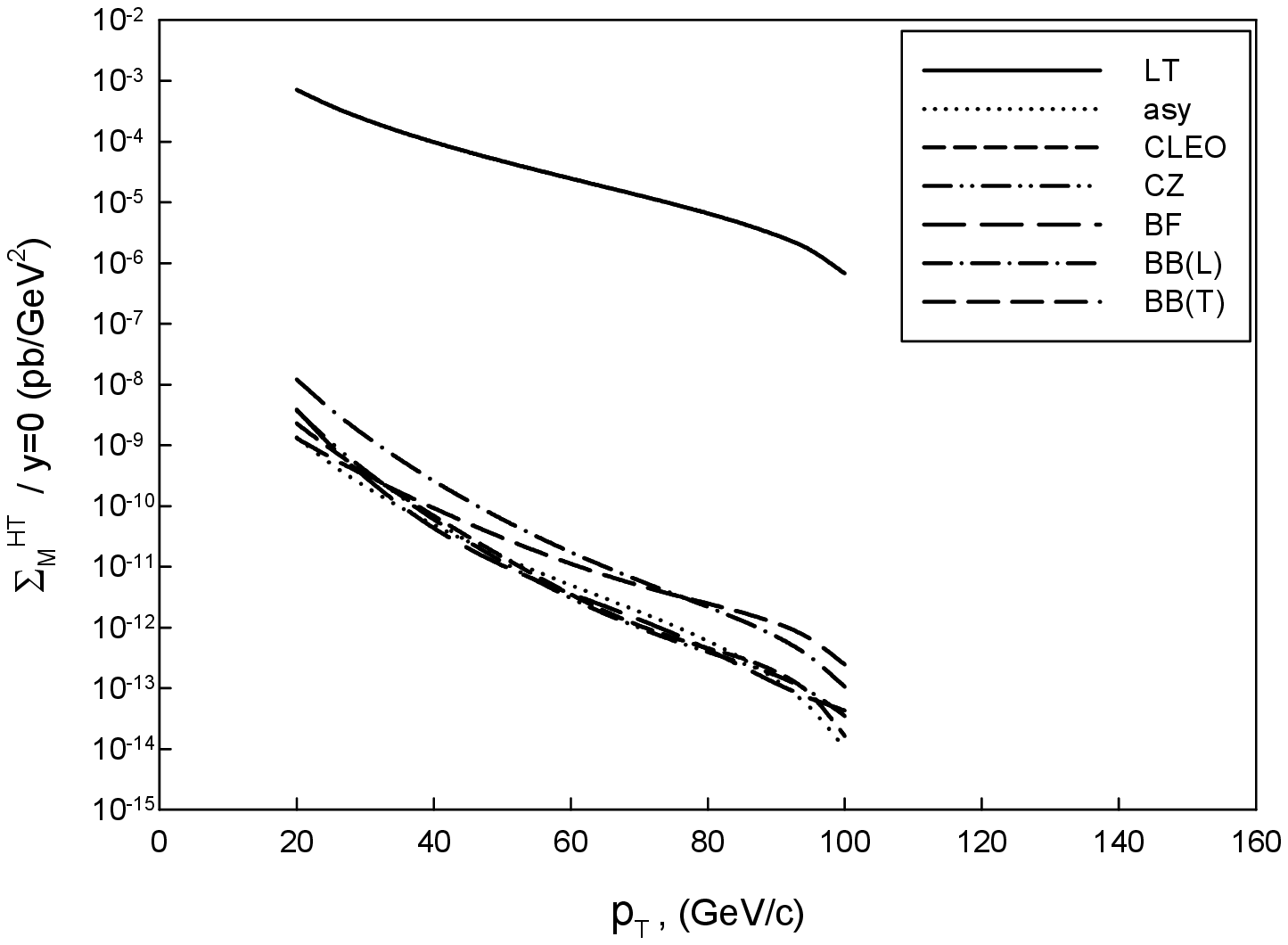}}\vskip-17.6cm \caption{High twist
meson $M$ production cross sections as a function of the $p_T$
transverse momentum of the meson at the c.m.energy $\sqrt s=209
\,\,GeV$.} \label{Fig5}
\end{figure}

\newpage

\begin{figure}[htb]
\vskip-2.1cm \epsfxsize 19.3cm \centerline{\epsfbox{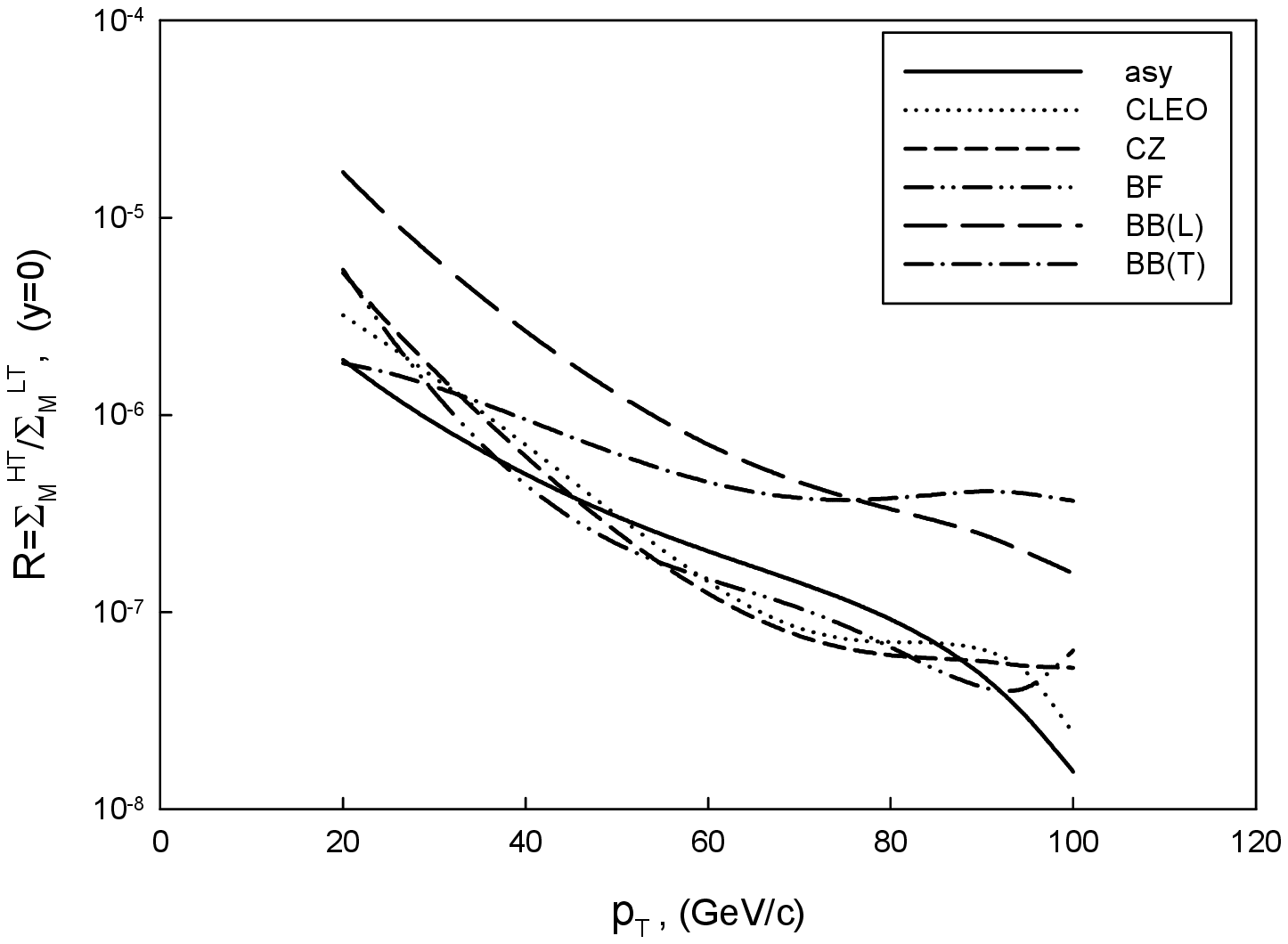}}
\vskip-17.7cm \caption{Ratio
$R=\Sigma_{M}^{HT}$/$\Sigma_{M}^{LT}$, where the leading and the
high twist contributions are calculated for the meson rapidity
$y=0$ at the c.m. energy $\sqrt s=209 \,\,GeV$, as a function of
the meson transverse momentum, $p_T$.} \label{Fig6}
\end{figure}
 \vskip-0.0cm

\begin{figure}[htb]
\vskip-1.7cm \epsfxsize 18.5cm \centerline{\epsfbox{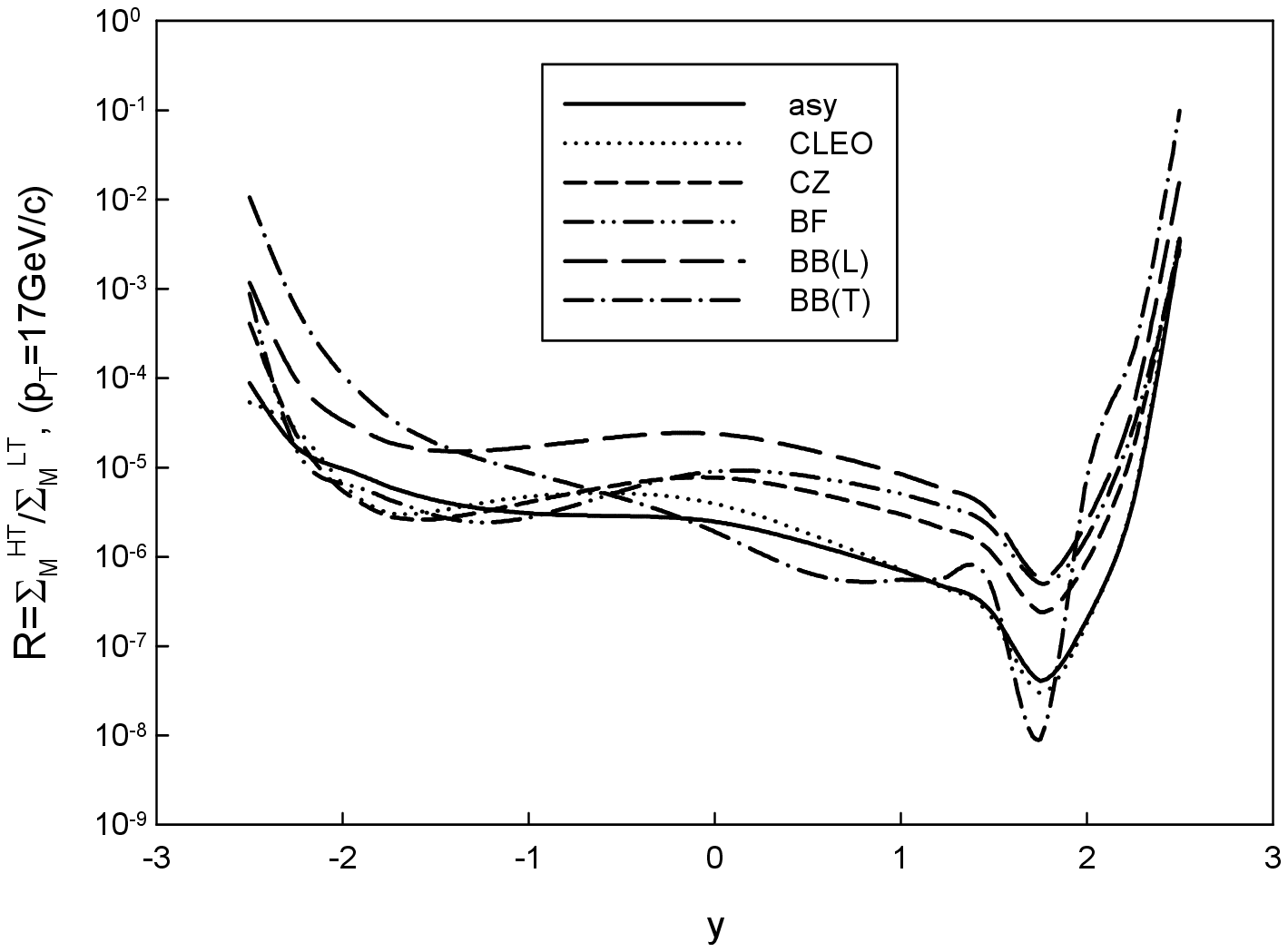}}
\vskip-17.2cm \caption{Ratio
$R=\Sigma_{M}^{HT}$/$\Sigma_{M}^{LT}$, where the leading and the
high twist contributions are calculated for the meson transverse
momentum $p_T=17 GeV/c$ at the c.m. energy $\sqrt s=209 \,\,GeV$,
as a function of the $y$ rapidity of the meson.} \label{Fig7}
\end{figure}


\begin{thebibliography}{99}
\section{References}
\bibitem{1}A.V.~Radyushkin, Dubna preprint P2-10717, 1977,
hep-ph/0410276.
\bibitem{2} V. L.~Chernyak and A. R.~Zhitnitsky, Nucl. Phys.
\textbf{B201}, 492 (1982); V. L.~Chernyak, A. R.~Zhitnitsky and I.
R.~Zhitnitsky, Nucl. Phys. \textbf{B204}, 477 (1982).
\bibitem{3} V. L.~Chernyak and A. R.~Zhitnitsky, Nucl. Phys.
\textbf{B246}, 52 (1984).
\bibitem{4} I. D.~King, C. T.~Sachrajda, Nucl. Phys. \textbf{B279},
785 (1987).
\bibitem{5} V. L.~Chernyak, A. R.~Zhitnitsky, Phys. Rept. \textbf{112},
173 (1984).
\bibitem{6} S. V.~Mikhailov and A. V.~Radyushkin, JETP Lett.
\textbf{43}, 712 (1986); Sov. J. Nucl. Phys.\textbf{49}, 494
(1989); Phys. Rev. \textbf{D45}, 1754 (1992).
\bibitem{7}V. M.~Braun and I. E.~Filyanov, Z. Phys. \textbf{C44}, 157
(1989).
\bibitem{8} G. R.~Farrar, K.~Huleihel and H.~Zhang, Nucl. Phys.
\textbf{B349}, 655 (1991).
\bibitem{9} V. L.~Chernyak, A. A.~Ogloblin and I. R.~Zhitnitsky, Z.
Phys. \textbf{C42}, 583  (1989).
\bibitem{10} G. R.~Farrar, H.~Zhang, A. A.~Ogloblin and I.
R.~Zhitnitsky, Nucl. Phys. \textbf{B311}, 585 (1988).
\bibitem{11} M.~Anselmino, P.~Kroll and B.~Pire, Z. Phys. \textbf{C36},
89 (1987).
\bibitem{12}A. V.~Radyushkin and R.~Ruskov, Phys. Lett. \textbf{B374},
173 (1996); Nucl. Phys. \textbf{B481}, 625 (1996).
\bibitem{13}S. J.~Brodsky and G. L.~Lepage, in: Perturbative Quantum
  Chromodynamics, ed. by A. H.~Mueller, p.93, World Scientific (Singapore) 1989.
\bibitem{14}S. J.~Brodsky, H.-C.Pauli and S. S.~Pinsky, Phys. Rept. \textbf{301},
  299 (1998).
\bibitem{15} The CLEO Collaboration (J.Gronberg
$et.al.$), Phys. Rev. \textbf{D57}, 33 (1998).
\bibitem{16} V. Yu.~Petrov, M. V.~Polyakov, R.~Ruskov, C.~Weiss and
K.~Goeke, Phys. Rev. \textbf{D59}, 114018  (1999); hep-ph/9807229.
\bibitem{17}J.~Qiu and G.~Sterman, Nucl. Phys. \textbf{B353}, 105, 137
(1991).
\bibitem{18}S. J.~Brodsky, Int. J. Mod. Phys. \textbf{A13}, 2417 (1998).
\bibitem{19}J. A.~Bagger and J. F.~Gunion, Phys. Rev. \textbf{D29}, 40 (1984).
\bibitem{20}J. A.~Bagger and J. F.~Gunion, Phys. Rev. \textbf{D25}, 2287 (1982).
\bibitem{21}J. A.~Hassan and J. K.~Storrow, Z. Phys. C-Particles and Fields 14, 65
(1982).
\bibitem{22}S. J.~Brodsky, T. A.~DeGrand, J. F.~Gunion and
J. H.~Weis, Phys. Rev. Lett \textbf{41}, 672 (1978); Phys. Rev. \textbf{D19},
1418 (1979).
\bibitem{23} G. P.~Lepage and S. J. ~Brodsky, Phys. Lett. \textbf{B87}, 359
  (1979); Phys. Rev. Lett. \textbf{43}, 545 (1979); 43, 1625(E) (1979);
 A.~Duncan and A.~Mueller, Phys. Rev. \textbf{D21}, 1636 (1980).
\bibitem{24}G. P.~Lepage and S. J.~Brodsky, Phys. Rev. \textbf{D22},  2157 (1980).
\bibitem{25} A. I.~Ahmadov, I.~Boztosun, R. Kh.~Muradov, A.~Soylu
and E. A.~Dadashov, Int. J. Mod. Phys. \textbf{E15}, 1209 (2006);
hep-ph/0607238.
\bibitem{26}E. L.~Berger and S. J. Brodsky,
Phys. Rev. Lett. \textbf{42}, 940 (1979); E. L. Berger, Z. Phys.
\textbf{C4}, 289 (1980).
\bibitem{27}V. N.~Baier and A. G.~Grozin, Phys. Lett. \textbf{B96}, 181
  (1980); S.~Gupta, Phys. Rev. \textbf{D24}, 1169 (1981).
\bibitem{28} A.~Schmedding and O. Yakovlev, Phys. Rev. \textbf{D62}, 116002
  (2000), hep-ph/9905392.
\bibitem{29}P.~Ball and V. M.~Braun, Phys. Rev. \textbf{D54}, 2182 (1996); hep-ph/9602323.
\bibitem{30} V. L.~Chernyak, Preprint TPI-MINN-91-47-T, December 1991.
\bibitem{31} M. A.~Shifman, A. I.~Vainshtein and  V. I.~Zakharov, Nucl. Phys. \textbf{B147}, 385 
(1979).
\bibitem{32} F. Cornet, Acta Phys. Polon. \textbf{B37}, 663 (2006); hep-ph/0601056.
\bibitem{33} B. A.~Kniehl, G.~Kramer, B.~P\"otter, Nucl.Phys. \textbf{B582}, 514 (2000), 
hep-ph/0010289.


\end{thebibliography}
\end{document}